\documentclass[%
 aip,
 jmp,%
 amsmath,amssymb,
 reprint,%
]{revtex4-1}

\usepackage{graphicx}
\usepackage{dcolumn}
\usepackage{bm}

\begin{document}

\preprint{AIP/123-QED}

\title{Testing the variation of the fine structure constant with strongly lensed gravitational waves}

\author{Xin Li}
\email{lixin1981@cqu.edu.cn}
 \affiliation{Department of Physics, Chongqing University, Chongqing 401331, China}
\author{Li Tang}%
 \email{tang@cqu.edu.cn}
\affiliation{Department of Physics, Chongqing University, Chongqing 401331, China}
\author{Hai-Nan Lin}
 \email{linhn@ihep.ac.cn}
\affiliation{Department of Physics, Chongqing University, Chongqing 401331, China}
\author{Li-Li Wang}
 \email{20152702016@cqu.edu.cn}
\affiliation{Department of Physics, Chongqing University, Chongqing 401331, China}
 %


\begin{abstract}
The possible variation of the electromagnetic fine structure constant $\alpha_e$ on cosmological scales arouses great interests in recent years. The strongly lensed gravitational waves and the electromagnetic counterparts could be used to test this variation. Under the assumption that the speed of photon could be modified, while the speed of GW is the same as GR predicated, and they both propagate in a flat Friedman-Robertson-Walker universe, we investigate the difference of time delays of the images and derive the upper bound of the variation of $\alpha_e$. For a typical lensing system in the standard cosmological models, we obtain $B\cos\theta\leq 1.85\times10^{-5}$, where $B$ is the dipolar amplitude and $\theta$ is the angle between observation and the preferred direction. Our result is consistent with the most up-to-date observations on $\alpha_e$. In addition, the observations of strongly lensed gravitational waves and the electromagnetic counterparts could be used to test which types of alternative theories of gravity can account for the variation of $\alpha_e$.

\end{abstract}

\maketitle

\section{Introduction}
Gravitational waves (GW), as one of the predictions of general relativity, has been detected recently by the advanced LIGO detector \cite{Abbott et al.:2016a}. Up to now, the LIGO and Virgo Collaborations have directly observed five GW events produced by the mergers of compact binary systems \cite{Abbott et al.:2016a,Abbott et al.:2016b,Abbott et al.:2017a,Abbott et al.:2017b,Abbott et al.:2017c}. The first four events were produced by the merge of binary black hole systems. The last one, GW170817, was produced by the merge of binary neutron star system, and the corresponding electromagnetic (EM) counterparts have been detected by many instruments \cite{EM1,EM2,EM3,EM4,EM5,EM6,EM7}. The observations of GW can be used to test cosmology and general relativity. One important cosmological quantity, i.e. the luminosity distance of the source, can be derived directly from the GW signal. Location of the source can be found from EM counterparts and the redshift of the source can be found from the association of the source with its host galaxy.

These information obtained from the observations of GW can be used to constrain the cosmological parameters, such as the equation-of-state of dark energy, the Hubble constant, etc \cite{Schutz:1986,Markovic:1993,Dalal:2006,Taylor:2012,Guidorzi:2017ogy}. Also, the GW signal has been used to constrain the mass of graviton \cite{Abbott et al.:2016a}, and the relative arriving time between the GW signals of GW170817 and its EM counterparts has been used to constrain the Lorentz invariance violation \cite{Will:2014,Abbott et al.:2016c,Liu:2016edq,Wu et al.:2016,Kahya and Desai:2016}. However, the intrinsic time delay in the emission time of GW signal and its EM counterpart can not be measured directly. To test the Lorentz invariance violation more precisely, an approach using the strongly lensed GW is proposed to cancel the intrinsic time delay \cite{Biesiada and Piorkowska:2009,Fan et al.:2017,Collett and Bacon:2017}. This approach requires GW and its EM counterpart occur behind a strong gravitational lensing and two images are observed.

Up to now, such phenomena has not yet been observed by the astronomical instruments. The LIGO and Virgo collaborations established a program\setcounter{footnote}{1}\footnote{http://www.ligo.org/scientists/GWEMalerts.php.} about the identification and follow-up of EM counterparts, which activates the campaign to find EM counterparts \cite{Abadie et al.:2012,Evans et al.:2012,Aasi et al.:2014}. The on-going third generation detectors with higher sensitivity such as the Einstein Telescope \cite{Einstein Telescope} will discover more GW events \cite{Biesiada et al.:2014,Ding et al.:2015}. The plausibility of such phenomena being observed is discussed in Ref.\cite{Collett and Bacon:2017}. It is expected that GW and its EM counterpart could be observed behind a strong gravitational lensing in the future.

Testing the constancy of the fundamental physical constants is very important \cite{Uzan}. Analyzing from observations of quasar absorption spectra, Webb et al. \cite{Webb et al.:2011,King et al.:2012} found that the fine structure constant $\alpha_e$ varies at cosmological scales. However, debates still remain \cite{Levshakov,Whitmore}. The GW signals provide us a new window to study the variation of $\alpha_e$. It is very interesting to test the variation of $\alpha_e$ by the observations of GW signals and its EM counterparts. Many models have been proposed to explain the variation of $\alpha_e$. These models can be divided into two types. The first is that the EM field is coupled to other field, such as quintessence field \cite{Copeland,Martins}. Other is that our universe is anisotropic, for example, our universe is a Finsler spacetime instead of Riemann spacetime \cite{Li et al.:2015,Li and Lin:2017}. If $\alpha_e$ does vary at cosmological scales, then the observations of the strongly lensed GW signals and their EM counterparts could be used to test which type of model is valid. This is due to the fact that the method proposed by Refs.\cite{Fan et al.:2017,Collett and Bacon:2017} mainly considers the difference between the time delay of two images of the GW signals and their EM counterparts. The first type of models requires that the speeds of photon and GW are different. Then the observations of the strongly lensed GW signals and their electromagnetic counterparts will find the difference from the method \cite{Fan et al.:2017,Collett and Bacon:2017}. The other type of model requires that both photon and GW propagate in the anisotropic universe with the same anisotropic speed. Then such observations will not find their difference. In this paper, we will discuss these points and show that the observations of the strongly lensed GW signals and their EM counterparts could test whether Webb's result is valid or invalid.

The arrangement of the paper is as follows: In Section 2, we introduce the basic information of GW and EM counterpart, and the method to calculate the difference of time delays between the GW and EM in detail. Then in Section 3 we use such method to constrain the variation of $\alpha_e$ and compare it with the observational data. Finally, conclusions and remarks are given in Section 4.

\section{Methodology}\label{sec:Methodology}

Webb et al. \cite{Webb et al.:2011,King et al.:2012} showed that the variation of fine structure constant $\alpha_e$ have a dipolar structure in high redshift region ($z>1.6$). Recently Pinho et al.\cite{Pinho and Martins:2016} also confirmed that the dipolar variation of $\alpha_e$ is still a good fit to the most up-to-date data. According to their results, the variation of $\alpha_e$ can be expressed as
\begin{equation}\label{formula alpha}
\frac{\triangle\alpha_e}{\alpha_e}=B\cos\theta,
\end{equation}
where $B$ represents the dipole amplitude and is assumed to be a constant, and $\theta$ is the angle between the dipolar direction and the observed direction.

One direct reason of the variation of $\alpha_e$ is the variation of speed of light. Therefore, from the (\ref{formula alpha}), the speed of light $c$ is anisotropic with dipolar structure in the cosmological scale,
\begin{equation}\label{c_gamma}
c_\gamma=c_0/(1+B\cos\theta),
\end{equation}
where $c_0$ is the speed of light at present epoch.

The method proposed by Refs.\cite{Fan et al.:2017,Collett and Bacon:2017} considers GW and its EM counterparts propagating through a strong gravitational lensing and at least two images of the GW event and two images of the EM counterparts are observed. The observations of this phenomena can detect two arriving time of GW events and two arriving time of EM events. The time delay of the two GW events does not depend on the initial emission time of the GW signals. The time delay of the two EM events also does not depend on the initial emission time of the EM signals. Thus, this method does not depend on the intrinsic separation time of GW and its EM counterparts.

In general relativity, the two time delay should be the same. If the difference of the two time delay is observed, then it is a signal of new physics. Two reasons will deduce the difference of the two time delay. One reason is that the speeds of graviton and photon are different. The other is that the geodesics of GW and photon are different. One type of model, such as Refs.\cite{Li et al.:2015,Li and Lin:2017}, could explain Webb's results by assuming the photon propagate in a Finslerian universe. In such model, both the graviton and photon are massless and their geodesic are the same. Therefore, if Webb's results are confirmed in future astronomical observations and the observations show no difference between the two time delay, then it implies our universe may be Finslerian.
Webb's results could be explained by another types of model which assume the speed of photon is modified. In alternative theories of gravity, both the speed of photon and graviton could be modified. Several approaches could lead to the modifications. For example, the speed of photon could be modified if Lagrange of electromagnetic field possesses the non-minimal coupling form \cite{Copeland,Martins}. And the speed of gravity could be modified if graviton couple to background gravitational field, such as massive gravity \cite{de Rham,Hinterbichler}. In Refs.\cite{Copeland,Martins}, the graviton or scalar curvature in Lagrange does not couple to background gravitational field, thus, the speed of graviton is unchanged in these models. Due to that the speeds of light and graviton are different, the difference between the two time delay should occurs. In the rest of our paper, we will mainly discuss how to test Webb's results with a model-independent method, which is only based on the assumption that the speed of light is different with the graviton.

The spatial geometry of FRW spacetime would not greatly affect the difference between the two time delay. And the recent data of Planck satellite prefers a spatially flat universe\cite{Planck XIII}. For simplicity, we suppose the spacetime is depicted by the flat FRW metric
\begin{equation}
ds^2=c^2dt^2-a(t)\left[dr^2+r^2d\theta^2+r^2\sin^2\theta d\phi^2\right],
\end{equation}
where $a(t)$ is the scale factor. Hence the travel distance of the GW from the emitted moment $t_e$ to observed moment $t_0$ is
\begin{equation}\label{rGW}
r_{GW}=\int_{t_e}^{t_0}\frac{c_0}{a(t)}dt=\frac{c_0}{H_0}\tilde{r}_{GW}(z),
\end{equation}
where $H_0$ is the Hubble constant, and
\begin{equation}
\tilde{r}_{GW}(z)=\int_{0}^{z}\frac{dz'}{E(z')}
\end{equation}
is the reduced comoving distance travelled by the graviton, where $E(z)=\sqrt{\Omega_{m}(1+z)^3+(1-\Omega_{m})}$. As for EM counterpart, the photon travels with the speed $c_{\gamma}$ given in Eq.(\ref{c_gamma}) and the corresponding distance is given as
\begin{equation}\label{r_gamma}
r_{\gamma}=\int_{t_e}^{t_0}\frac{c_{\gamma}}{a(t)}dt=\frac{c_0}{H_0}\tilde{r}_{\gamma},
\end{equation}
where $\tilde{r}_{\gamma}=\left[1-f(B,\theta)\right]\tilde{r}_{GW}$ and $f(B,\theta)=B\cos\theta-B^2\cos^2\theta$. Here, we have use the fact that the magnitude $B$ of $\alpha_e$ variation is very small and expand $f(B,\theta)$ to the second order of $B$.

In a strong gravitational lensing system, the time delay between two collinear images which are observed on the opposite side of the lens is given as
\begin{equation}\label{delta t GW}
\Delta t=\frac{1+z_l}{2c}\frac{D_lD_s}{D_{ls}}(\theta^2_A-\theta^2_B),
\end{equation}
where $\theta_A=\theta_E+\beta$ and $\theta_B=\theta_E-\beta$ are the radial distances of two images, respectively, and $\beta$ denotes the misalignment angle. Here, $D_{ls}$ denotes the angular diameter distance between the lens and source, and $D_l$ ($D_s$) denotes the angular diameter distance between the lens  (source) and observer. In the singular isothermal sphere lens model, the Einstein ring radius takes the form
\begin{equation}\label{theta_E}
\theta_E=4\pi\frac{D_{ls}}{D_s}\frac{\sigma^2}{c^2},
\end{equation}
where $\sigma$ is the one dimensional velocity dispersion. Combining eqs.(\ref{rGW})(\ref{delta t GW})(\ref{theta_E}), the time delays of two GW signals is given as
\begin{equation}\label{eq:delta_t}
\Delta t_{GW}=\frac{32\pi^2}{H_0}\left(\frac{\sigma}{c_0}\right)^4\frac{\beta\tilde{r}(z_l)\tilde{r}(z_l,z_s)}{\theta_E\tilde{r}(z_s)}.
\end{equation}

The electromagnetic field with a non-minimal coupling, such as refs.\cite{Copeland,Martins} implies the photon is massive. From the geodesic equation in general relativity, one can find that the deflection of the photon with a small rest mass would be altered with a factor $1+(m^2_{\gamma}c_0^4/2E^2_{\gamma})$, where $m_{\gamma}$ and $E_{\gamma}$ are the mass and energy of the photon. Thus, the Einstein radius is modified as $\theta_{E,\gamma}=\theta_E[1+(m^2_{\gamma}c_0^4/2E_{\gamma}^2)]$ and the time delay between the two images of EM is given as
\begin{equation}\label{delta t gamma}
\Delta t_{\gamma}=\frac{32\pi^2}{H_0}\left(\frac{\sigma}{c_0}\right)^4\frac{\beta\tilde{r}_{\gamma}(z_l) \tilde{r}_{\gamma}(z_l,z_s)}{\theta_E\tilde{r}_{\gamma}(z_s)}\left[1+\frac{m^2_{\gamma}c_0^4}{2E_{\gamma}^2}\right].
\end{equation}
In flat FRW spacetime, the spacetime is Minkowski spacetime in local. Thus, the dispersion relation of massive photon is the same with other massive particles in Minkowski spacetime. Combining the eq.(\ref{c_gamma}) and the dispersion relation, the relation between the mass of the photon and its speed is derived as
\begin{equation}\label{m-v}
\frac{m^2_{\gamma}c^4_0}{2E_{\gamma}^2}=\frac{1}{2}\left(1-\frac{1}{(1+B\cos\theta)^2}\right).
\end{equation}
By making use of the formulae (\ref{delta t GW},\ref{delta t gamma}) and eq.(\ref{m-v}), to second order in $B$, we obtain the difference of the two time delay
\begin{equation}\label{band}
\Delta t_{GW}-\Delta t_{\gamma}=\Delta t_{GW}\frac{3}{2}B^2\cos^2\theta.
\end{equation}

\section{Results}\label{sec:Result}

The observational accuracy $\delta T$ of the difference between two time delays, i.e., $\Delta t_{GW}-\Delta t_{\gamma}$, could give a constraint on the dipole variation of $\alpha_e$. From the eq.(\ref{band}), we find that
\begin{equation}
B\cos\theta\leq\left(\frac{2}{3}\frac{\delta T}{\Delta t_{GW}}\right)^{1/2}.
\end{equation}
In the strong gravitational lensing systems compiled in Ref. \cite{Cao et al.:2015}, the redshift ranges are $z_l$ $\in$ $[0.075,1.004]$ for the lens and $z_s$ $\in$ $[0.196,3.596]$ for the source, and the velocity dispersions are at the range $\sigma$ $\in$ $[103,391]$ km/s. Additionally, the source-lens misalignment parameter $\beta/\theta_E$ should not be to large in order to ensure the formation of multiple images. Pi\'{o}rkowska et al. \cite{Piorkowska et al.:2013} demonstrated that the maximal value of misalignment parameter $\beta/\theta_E$ is $0.5$.
For the timing accuracy $\delta T$ of time delay, observation has demonstrated that the GW signal can be detected at precision $<10^{-4} {\rm ms}$ \cite{Abbott et al.:2016a,Fan et al.:2017}. Moreover, the timing precision of promising EM counterparts, such as SGRB and FRB, could be in the order of 10$^{-2}-10^3$ms \cite{Fox et al.:2005,Champion et al.:2016}. Thus, the accuracy of the EM time delay determine the ability of testing Webb's result. However, since the strongly lensed gravitational waves and their electromagnetic counterparts have not been detected,  $\delta T=1$ms could be set as a mediate timing precision of promising EM counterparts to obtain the detecting precision of testing the $\alpha_e$ variation.


Considering the $\Lambda$CDM cosmology parameters given by the Planck data\cite{Planck XIII}, i.e., $H_0=68~\rm km s^{-1} Mpc^{-1}$, $\Omega_{M0} = 0.3$, and using the typical parameters of strong lensing system ($z_l=1$, $z_s=2$, $\sigma=250~\rm km/s$, and $\beta/\theta_E=0.1$), and assuming the timing accuracy $\delta T=1$ ms, we obtain the bound of $\alpha_e$ variation as
\begin{equation}\label{Bcos}
B\cos\theta\leq 1.85 \times10^{-5}.
\end{equation}
Webb et al. \cite{Webb et al.:2011} showed that the magnitude of $\alpha_e$ variation is $(0.97^{+0.22}_{-0.20})\times 10^{-5}$. Pinho et al. \cite{Pinho and Martins:2016} showed that the magnitude of $\alpha_e$ variation is $(0.81\pm0.17)\times 10^{-5}$. Thus, in the detecting precision, the constraint on $\alpha_e$ variation eq.(\ref{Bcos}) is consistent with the previous researches on the $\alpha_e$ variation. This implies that the observations of the difference between the GW time delay and EM time delay are capable of testing whether the $\alpha_e$ variation is valid or not.

The upper limit of the variation of $\alpha_e$ measured in the Milky Way is $|\Delta\alpha_e/\alpha_e|<1.1\times 10^{-7}$ \cite{Levshakov:2009}. Thus, if Webb' result is correct, there should be a physical mechanism that $\alpha_e$ varies with redshift. In fact, our previous research has shown one such possible physical mechanism \cite{Li and Lin:2017}, where the speed of light depends on the redshift with the form
\begin{equation}\label{c_gamma2}
c_\gamma=c_0/(1+B(z)\cos\theta),
\end{equation}
where
\begin{equation}\label{Bz}
B(z)=b_0\int^{z}_{0}\frac{1+z^{'}}{\sqrt{\Omega_{m}(1+z^{'})^3+(1-\Omega_{m})}}dz^{'}=b_0D(z).
\end{equation}
here $D(z)=\int^z_0\frac{1+z^{'}}{E(z')}dz^{'}$. To the first order of $b_0$, the difference of time delays measured by the GW and EM windows becomes
\begin{equation}\label{band2}
\Delta t_{GW}-\Delta t_{\gamma}=\Delta t_{GW}b_0\cos\theta F(z_l,z_s),
\end{equation}
where
\begin{equation}
F=\frac{x(z_s)}{\tilde{r}_{g,ls}}-\frac{x(z_l)}{\tilde{r}_{g,ls}}+\frac{x(z_l)}{\tilde{r}_{g,l}}-\frac{x(z_s)}{\tilde{r}_{g,s}}-D(z_s)
\end{equation}
here $x(z)=\int^z_0\frac{D(z')}{E(z')}dz'$.
It is different with eq.(\ref{band}), where the formula represents that the difference of time delays is proportion to $B^2$. As eq.(\ref{Bz}) is shown that, if the variation of $\alpha_e$ is independent on the redshift, viz, $D(z)=1$ in eq.(\ref{Bz}), the term which is proportional to $B$ would vanish in eq.(\ref{band2})($F$=0). Then, the formula (\ref{band2}) reduces to the formula (\ref{band}) under the consideration of the second order of $B$. With the same variable setting as above that $z_s=2$ and $z_l=1$, the upper bound of dipolar variation $b_0\cos\theta \leq2.08\times 10^{-10}$, which reaches a very high accuracy to test the variation of the fine structure constant.

It should be noted that Webb's result about the dipolar variation of $\alpha_e$ mainly appears in high redshift region ($z>1.6$). Furthermore, our method, i.e., testing the variation using the difference between the GW time delay and EM time delay, needs the observation of EM counterparts of GW signals. However, due to the present sensitivity of LIGO detector, it is incapable to detect GW signals with EM counterparts locating at redshift $z>1$. The on-going third generation detectors like the Einstein Telescope \cite{Einstein Telescope} with higher sensitivity are capable of testing Webb's result.

\section{Conclusions and Remarks}\label{sec:summary}

The associated detection of GWs and their EM counterparts provides a way to test fundamental physics. In this paper, we have used the method proposed by Refs.\cite{Fan et al.:2017,Collett and Bacon:2017} to test the possible variation of $\alpha_e$. The method considers the difference between the time delay of two images of the GW event and its EM counterparts. The difference between the speeds of photon and graviton, and the difference between the geodesic of photon and graviton, can account for the difference between the two time delay. In an anisotropic universe, the geodesic and speeds of photon and graviton are modified in the same way. Therefore, if Webb's results are confirmed by the future data and the observations show no difference between the two time delay, then it implies our universe may be anisotropic, such as Finslerian universe.

In this paper, we consider that the speed of photon is modified, such as electromagnetic field coupling to a quintessence field\cite{Copeland,Martins}. In these models the speed of graviton remains the same as the prediction of general relativity, and both the graviton and photon propagate in the same flat FRW universe. It is shown that the dipolar variation of $\alpha_e$ has a upper limit, namely, $B\cos\theta\leq 1.85\times10^{-5}$, which implies that Webb's result can be tested in currently accuracy with this method. Additionally, considering the variation of $\alpha_e$ could be a function of the redshift as Li $\&$ Lin \cite{Li and Lin:2017}, we obtained a bound of $\alpha_e$ variation, $b_0\cos\theta\leq 2.08\times10^{-10}$, which is a higher detecting precision to test the result of Webb.

One should notice, due to the present sensitivity of LIGO detector, one can not find GW signals with EM counterparts locating at redshift $z>1$. Since, Webb et al. found the dipolar variation of $\alpha_e$ only appear in high redshift ($z>1.6$) region. Thus, it is expected that the on-going third generation detectors such as the Einstein Telescope could test the validity of Webb's result by observing the difference between the time delay of two images of the GW event and its EM counterparts. In this paper, we only give a limit of $B\cos\theta$, the constraint on $B$ and the preferred direction are not given. If many events of the strongly lensed GWs and their EM counterparts are observed in the future, then it is possible to use the data to constrain the dipolar amplitude $B$ and the preferred direction of the universe.

\acknowledgments{We are grateful to H. Wen, S. P. Zhao and C. Ye for useful discussions. This work has been supported by the National Natural Science Fund of China (Grant Nos. 11775038, 11603005, and 11647307).}

\end{document}